\pdfoutput=1
\documentclass[aps,prd,twocolumn]{revtex4-1}

\usepackage[colorlinks=true, pdfstartview=FitV, linkcolor=blue, citecolor=blue, urlcolor=blue]{hyperref}
\usepackage[dvipdfmx]{graphicx}
\usepackage{amsmath}
\usepackage{color}

\newcommand\RIGHT{\mathrm{R}}
\newcommand\LEFT{\mathrm{L}}
\newcommand\bv[1]{\boldsymbol{#1}}
\newcommand\Tr{\mathop{\mathrm{Tr}}}

\begin{document}

\title{Lattice field theory with torsion}

\author{Shota~Imaki}
\author{Arata~Yamamoto}
\affiliation{Department of Physics, The University of Tokyo, Tokyo 113-0033, Japan}

\begin{abstract}
Inspired by the duality between gravity and defects in crystals, we study lattice field theory with torsion.
The torsion is realized by a line defect of a lattice, namely a dislocation.
As the first application,
we perform the numerical computation
for vector and axial currents induced by a screw dislocation.
This current generation is called the chiral torsional effect.
We also derive the analytical formula for the chiral torsional effect in the continuum limit.
\end{abstract}

\maketitle

\section{Introduction}

Quantum field theory in a curved space shows various intriguing phenomena.
Although we want to observe and confirm such gravity-induced phenomena, direct observation is not easy.
In solid-state physics, there is an interesting idea that gravitational effects can be mimicked by distorted lattices in crystals.
(See textbooks~\cite{Kleinert:1989ky,Blagojevic:2013xpa} and references therein.)
Since this emergent gravity is more controllable than genuine gravity, we will have a better chance for direct observation.
Lattice defects behave as sources of the emergent gravity: a disclination corresponds to curvature and a dislocation corresponds to torsion.
There are many proposals to study gravity-induced phenomena through these lattice defects~\cite{Mesaros:2009az,deJuan:2010zz,Randono:2010cd,Gromov:2014vla,Zubkov:2015cba,Cortijo:2015jja,Sepehri:2016nuv,Castro:2018iqt}.

Since lattice field theory is formulated on a lattice, we can apply the same idea to lattice field theory.
Introducing lattice defects to lattice field theory, we can simulate quantum field theory in a curved space.
There are two motivations for this attempt.
First, this is a non-perturbative framework to study gravity-induced quantum phenomena.
Lattice field theory is a powerful computational scheme beyond perturbation. 
The lattice field theory in a curved space was previously formulated by introducing metric tensor and vierbein to lattice action~\cite{Yamamoto:2013zwa,Yamamoto:2014vda,Villegas:2014dqa}.
The use of lattice defects is another formulation.
The new formulation allows us to simulate not only curvature but also torsion.
Second, this is the exact calculation for the emergent gravity in solid-state physics.
Since most lattice systems with defects are analytically unsolvable, analytical calculations are impossible without approximations.
Ab-initio numerical calculations are necessary to get exact results.

In this paper, we study lattice field theory with lattice defects, in particular, with an external dislocation.
In the continuum limit, the dislocation corresponds to torsion, i.e., a space with non-commutative derivatives.
In Sec.~\ref{secTD}, we start with a brief introduction of torsion, and then discuss how to introduce dislocations to lattice field theory.
As the first application, we numerically compute the electric currents induced by torsion, which we call the ``chiral torsional effect'', in Sec.~\ref{secCTE}.
We also show the analytical formula of the chiral torsional effect in the continuum limit.
We discuss phenomenological applications in Sec.~\ref{secPA} and future perspectives in Sec.~\ref{secSP}.

\section{Torsion and dislocation}
\label{secTD}

In the Einstein gravity, the affine connection is assumed to be symmetric, $ \Gamma^\alpha_{\mu\nu} = \Gamma^\alpha_{\nu\mu}$.
In the Einstein-Cartan gravity, the antisymmetric part of the affine connection,
\begin{equation}
 T_{\mu\nu}^\alpha 
 = \frac12 \left( \Gamma^\alpha_{\mu\nu} - \Gamma^\alpha_{\nu\mu} \right),
\end{equation}
is nonzero.
This is called the torsion tensor.
In terms of the vierbein, the torsion tensor is given by
\begin{equation}
 T_{\mu\nu}^n 
 = \frac12 \left( \partial_\mu e_\nu^n - \partial_\nu e_\mu^n \right).
\end{equation}
Nonzero torsion implies the non-commutativity of derivatives, $[\partial_\mu, \partial_\nu] x^\alpha \neq 0$.
This expression clarifies that the nonzero torsion arises from the singular or multi-valued coordinate.
The non-commutative derivatives also appear in the study of topological vortices~\cite{Son:2004tq,Fukushima:2018ohd}.
The torsion couples to matter fields through the connection and the vierbein.
For example, the Dirac equation of a massless fermion reads
\begin{align}
	e^\nu_n \gamma^n D_\nu \psi = 0
\label{eqDirac}
\end{align}
with the covariant derivative
\begin{equation}
D_\nu = \partial_\nu + i\Gamma_\nu + \delta_{\nu\tau} (\mu + \gamma^5 \mu_5) .
\end{equation}
The Matsubara formalism with imaginary time $\tau$ is adopted.
The vector chemical potential $\mu$ and the axial chemical potential $\mu_5$ are introduced in the $\tau$ direction.

\begin{figure}[t]
\begin{center}
\includegraphics[width=.48\linewidth]{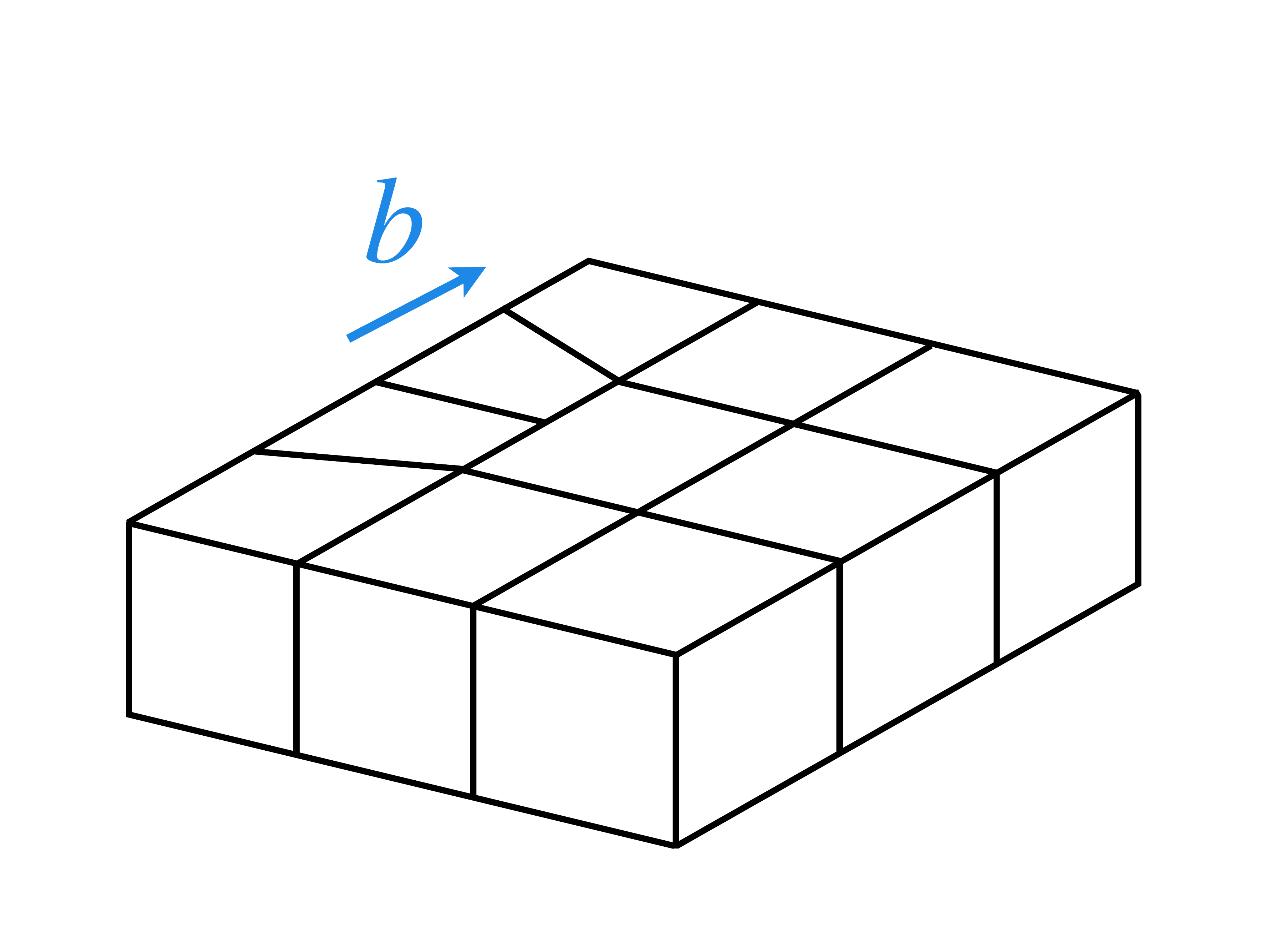}
\includegraphics[width=.48\linewidth]{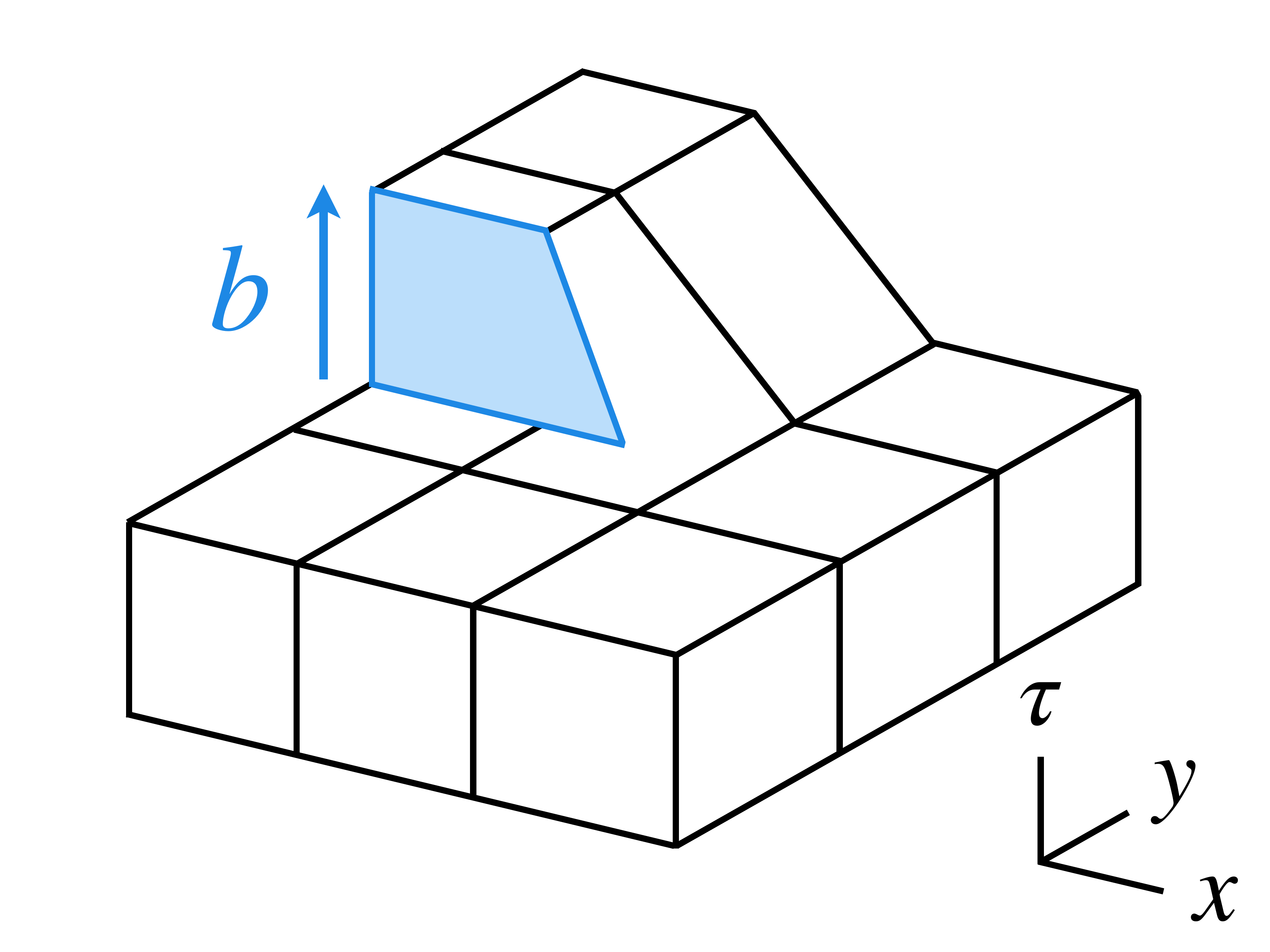}
\caption{
\label{fig}
Schematic figure of the edge dislocation (left) and the screw dislocation (right).
The arrows indicate the Burgers vectors.
}
\end{center}
\end{figure}

On the lattice, the torsion is realized by dislocations~\cite{Kleinert:1989ky}.
The schematic figure of dislocations is shown in Fig.~\ref{fig}.
The dislocation is parameterized by the displacement vector $b^\alpha$, which is called the Burgers vector.
The dislocation is classified into the edge one and the screw one in accordance with the direction of the Burgers vector.
In the continuum limit, the edge dislocation reproduces the torsion like $T^x_{xy}$ and the screw dislocation reproduces the torsion like $T^z_{xy}$.
In our formulation, we assume that the dislocations are externally fixed by hand.
Thus the corresponding torsion is external, not dynamical.

We consider formulating lattice field theory on the lattice with dislocations.
The calculation, except for lattice geometry, is the standard one in the lattice field theory in a flat space.
The standard lattice actions of scalar, fermion, and gauge fields can be used.
This is in contrast to the previous formulation of lattice field theory in a curved space~\cite{Yamamoto:2013zwa,Yamamoto:2014vda,Villegas:2014dqa}.
Dislocations do not break gauge invariance in gauge theory because the lattice still consists of the gauge-invariant loops of links.
Although the pure gauge theory with dislocations can be considered, it will not be interesting.
Link variables themselves do not distinguish directions unlike the Dirac operator, which explicitly depends on directions through the gamma matrices.
If one naively construct lattice gauge action from the loops of the link variables, it reproduces the continuum gauge action without torsion.
Therefore the dislocations do not have nontrivial effects on the pure gauge theory.
This is consistent with the fact that torsion does not couple to gauge fields in non-trivial and gauge-invariant manners~\cite{Shapiro:2001rz}.

\section{Chiral torsional effect}
\label{secCTE}

We first consider the Euclidean lattice with one screw dislocation.
The dislocation is placed at the center plaquette in the $x$-$y$ plane, as shown in the right panel of Fig.~\ref{fig}.
The Burgers vector is along the imaginary time direction, $b^\alpha = b \delta^{\alpha \tau}$.
In the continuum limit, this corresponds to the torsion tensor 
\begin{equation}
T_{xy}^\tau(x) = - \frac{b}{2} \delta^{(2)}(x) 
\label{eqdelta}
\end{equation}
with the two-dimensional delta function $\delta^{(2)}(x)$ in the $x$-$y$ plane.

\begin{figure*}[t]
\begin{center}
 \includegraphics[width=.48\textwidth]{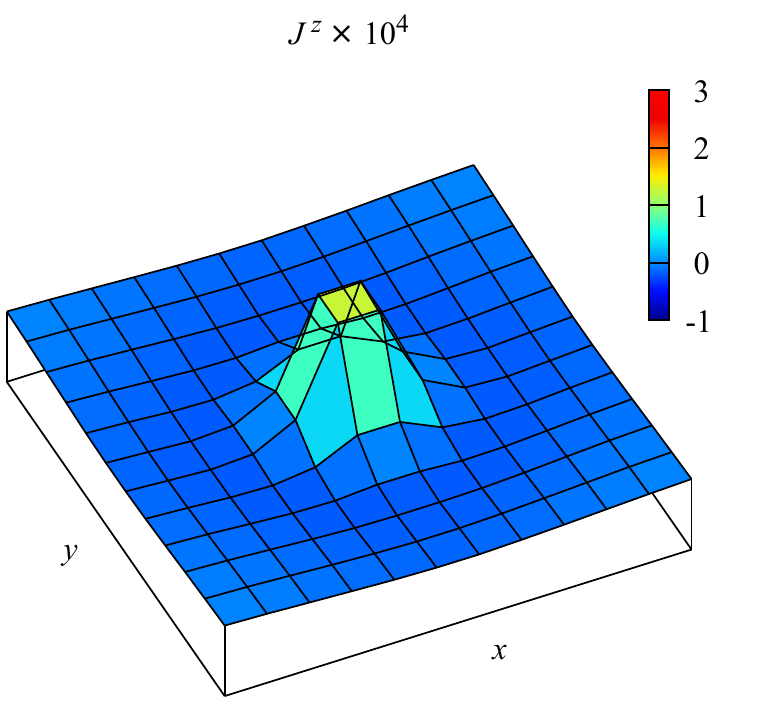}
 \includegraphics[width=.48\textwidth]{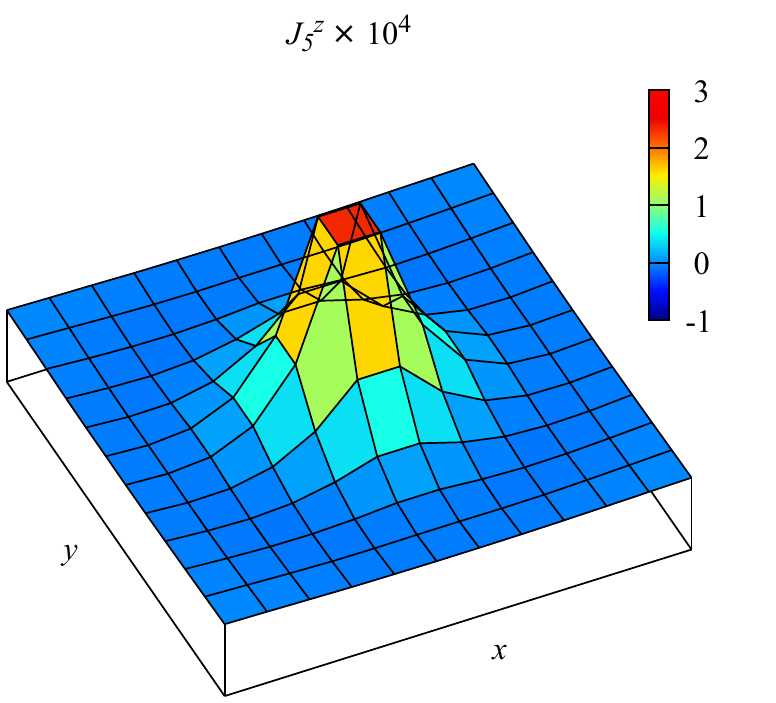}
\caption{
\label{figJ3D}
The vector current $J^z$ and the axial current $J_5^z$ in the $x$-$y$ plane.
The vector and axial chemical potentials are $\mu = \mu_5 = 0.1$, the temperature is $T = 1/12$, and the Burgers vector length is $b = 3$.
The lattice unit is used.
}
\end{center}
\end{figure*}

\begin{figure}[t]
\begin{center}
 \includegraphics[width=.48\textwidth]{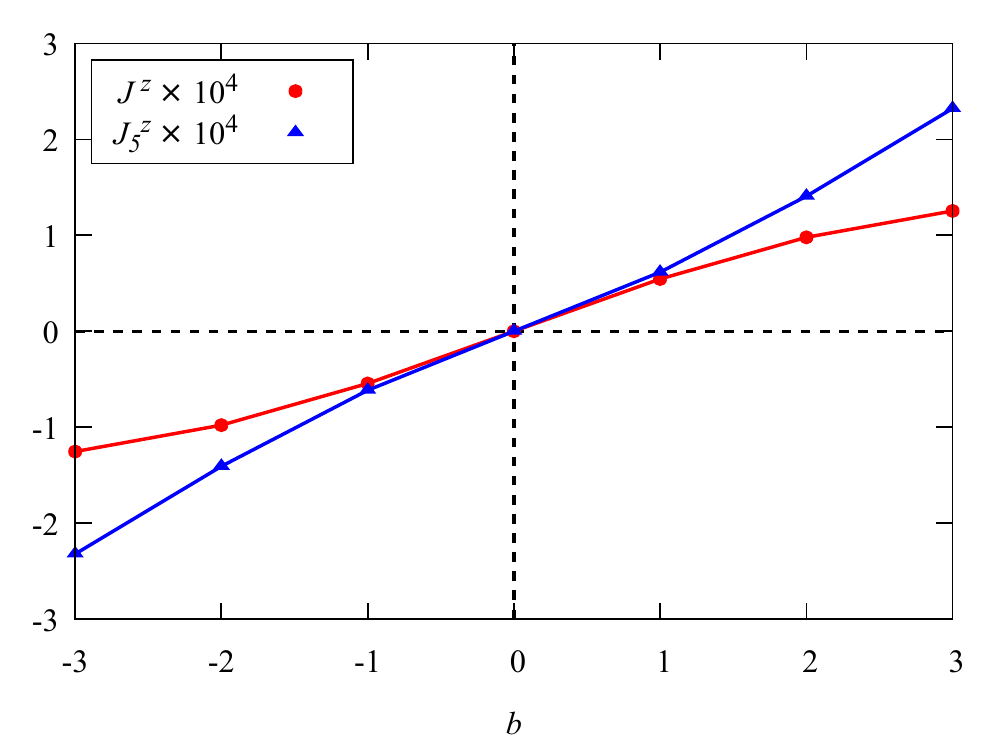}
\caption{
\label{figJ}
The peak values of the vector current $J^z$ and the axial current $J_5^z$ as functions of the Burgers vector length $b$.
The vector and axial chemical potentials are $\mu = \mu_5 = 0.1$ and the temperature is $T = 1/12$.
The lattice unit is used.
}
\end{center}
\end{figure}

We performed lattice calculations with a single massless Dirac fermion.
The fermion is coupled to vector and axial chemical potentials,
but not to dynamical gauge fields.
The details of the lattice calculations are given in Appendix \ref{secB}.
We calculated the vector current and the axial current in the $z$ direction.
As shown in Fig.~\ref{figJ3D}, the dislocation induces the currents.
The induced currents have peaks at the center plaquette.
In Fig.~\ref{figJ}, the peak values are plotted against the Burgers vector length $b$.
The currents are roughly proportional to $b$ while small non-linearity is also seen.

In the continuum and infinite-volume limit, we can derive the analytical expression of the vector current 
\begin{align}
	J^{z}
	= - \frac{\mu \mu_5}{\pi^2} T^\tau_{xy}
	\label{eqJV}
\end{align}
and the axial current
\begin{align}
	J_5^{z}
	= - \left( \frac{\mu^2 + \mu_5^2}{2 \pi^2} + \frac{T^2}{6} \right) T^\tau_{xy} ,
	\label{eqJA}
\end{align}
at the linear order of the torsion tensor.
The derivation is given in Appendix \ref{secA}.
The numerical results can be explained by Eqs.~\eqref{eqJV} and \eqref{eqJA}.
The peaks in Fig.~\ref{figJ3D} will reproduce the delta functions \eqref{eqdelta} in the continuum limit.
The peak values in Fig.~\ref{figJ} are proportional to $b$.
Although the peak values themselves cannot be directly compared with Eqs.~\eqref{eqJV} and \eqref{eqJA} because they are infinite due to the delta functions in the continuum limit, the ratio $J^{z}/J_5^{z}$ is comparable.
In this parameter setting, Eqs.~\eqref{eqJV} and \eqref{eqJA} give $J^{z}/J_5^{z} \simeq 1/2$.
This is consistent with the numerical results at $b=3$.
Note that large $b$ in the lattice unit corresponds to the continuum limit.

We call this current generation the ``chiral torsional effect'' 
\footnote{The name ``chiral torsional effect'' was used in Ref.~\cite{Khaidukov:2018oat} to call only the axial current \eqref{eqJA}.
We would like to use the same name for the vector current, too.}.
One can interpret the chiral torsional effect as an analog of the chiral vortical effect~\cite{Khaidukov:2018oat}.
The functional forms of Eqs.~\eqref{eqJV} and \eqref{eqJA} coincide with those of the chiral vortical effect by the replacement $T^\tau_{xy} \leftrightarrow -\omega_z$~\cite{Kharzeev:2015znc}.
We have shown the chiral torsional effect of a single free fermion.
The introduction of non-interacting multiple colors $N_c$ would give the overall factor $N_c$ to the currents.
The introduction of interactions would give more nontrivial change.

\section{Phenomenological application}
\label{secPA}

Although the chiral torsional effect is interesting from an academic viewpoint, we do not know how to make it in the real world.
The chiral torsional effect is induced by the torsion in the time direction.
It is extremely difficult to distort the temporal structure of matters.
Here we consider the torsion in spatial directions.
Unlike the torsion in the time direction, the torsion in spatial directions can be realized in realistic matters, e.g., in crystals with dislocations and in fluids with vorticities.
From Eqs.~\eqref{JV} and \eqref{JA}, the torsion $T^x_{zx}$ induces a vector current
\begin{align}
	J^{z}
	= \left( \frac{\mu^2 + \mu_5^2}{2 \pi^2} + \frac{T^2}{6} \right) T^x_{zx}
	\label{eqJVe}
\end{align}
and an axial current
\begin{align}
	J_{5}^z
	= \frac{\mu \mu_5}{\pi^2} T^x_{zx} .
	\label{eqJAe}
\end{align}
This is another pattern of the chiral torsional effect.
This can be studied by the lattice calculation with an edge dislocation.
Besides, the torsion $T^z_{xy}$ induces a chiral imbalance
\begin{equation}
\begin{split}
	n_{5} &= J_5^\tau
\\
	&= \frac{\mu_5^3}{3\pi^2} + \frac{\mu_5 \mu^2}{\pi^2} + \frac{\mu_5 T^2}{3}
	+ \left( \frac{\mu^2 + \mu_5^2}{2 \pi^2} + \frac{T^2}{6} \right) T^z_{xy} .
	\label{eqnA}
\end{split}
\end{equation}
The chiral imbalance is nonzero even at $\mu_5=0$.
This means that the torsion can stand in for the axial chemical potential.
We can also study many other torsion-induced and dislocation-induced phenomena found in earlier works~\cite{Mesaros:2009az,deJuan:2010zz,Randono:2010cd,Gromov:2014vla,Zubkov:2015cba,Cortijo:2015jja,Shitade:2013wda,Parrikar:2014usa,Sumiyoshi:2015eda,Chernodub:2015wxa,Ferreiros:2018udw}.

\section{Summary and perspective}
\label{secSP}

We proposed the idea to implement external gravity in lattice gauge theory by means of lattice defects.
In this work, a dislocation was introduced to realize torsion in the continuum limit.
We studied the chiral torsional effect.
Performing the lattice calculation with a dislocation, we observed the vector and axial currents of Dirac fermions.
The numerical results qualitatively agree with the analytical formulas in the continuum limit.
This idea can be generalized to other lattice defects, e.g., disclinations to realize curvature.

We performed the lattice calculation without interaction in this work.
Our formulation can be applied to the Monte Carlo simulation of interacting theory.
Since a dislocation only changes lattice structure and does not change lattice action, it seems free from the sign problem on the Monte Carlo simulation.
This is correct for the dislocations in spatial directions but incorrect for the dislocation in the real time direction.
The dislocation must be introduced in real time before the Wick rotation, not in imaginary time after the Wick rotation.
As a consequence, the dislocation in real time will make the lattice action complex, and thus cause the sign problem.
This is the same difficulty as the lattice gauge theory in a rotating frame~\cite{Yamamoto:2013zwa}.

\begin{acknowledgments}
S.~I.~was supported by Grant-in-Aid for JSPS Fellows Grant Number 19J22323.
A.~Y.~was supported by JSPS KAKENHI Grant Number 19K03841. 
The numerical calculations were carried out on SX-ACE in Osaka University.
\end{acknowledgments}

\appendix

\section{Lattice calculation}
\label{secB}

In this appendix, the detail of lattice calculation is explained.
The lattice calculation is formulated in the flat Euclidean space, and the index sum is explicitly written.
The lattice unit is assumed.

We used the non-interacting massless Wilson fermion.
The action is
\begin{equation}
\begin{split}
S =& \sum_x \sum_{\nu} \bar\psi(x) \bigg[ \psi(x) - \frac{1}{2} \bigg\{ (1-  \sum_{n} e_n^\nu \gamma^n U_{5\nu}) U_\nu \psi(x+\hat{\nu})
\\
& + (1+ \sum_{n} e_n^\nu \gamma^n U^{-1}_{5\nu}) U_\nu^{-1} \psi(x-\hat{\nu}) \bigg\} \bigg]
\end{split}
\end{equation}
with
\begin{equation}
\begin{split}
 U_{5\nu} &= \exp[+\mu_5\gamma^5 \delta_{\nu\tau}]
\\ 
 U^{-1}_{5\nu} &= \exp[-\mu_5\gamma^5 \delta_{\nu\tau}]
\end{split}
\end{equation}
and
\begin{equation}
\begin{split}
 U_\nu &= \exp[+\mu\delta_{\nu\tau}]
\\ 
 U^{-1}_\nu &= \exp[-\mu\delta_{\nu\tau}]
.
\end{split}
\end{equation}
The vector and axial chemical potentials are introduced on temporal link variables~\cite{Hasenfratz:1983ba,Yamamoto:2011gk}.
The vierbein is $e_1^x=e_2^y=e_3^z=e_4^\tau=1$, otherwise zero.
When the vector $\hat\nu$ crosses the cut in Fig.~\ref{fig}, it is given by
\begin{equation}
 \psi(x\pm\hat{y}) = \psi(x\pm\hat{2}\mp b\hat{4}),
\end{equation}
otherwise
\begin{equation}
\begin{split}
 \psi(x\pm\hat{x}) &= \psi(x\pm\hat{1})
\\ 
 \psi(x\pm\hat{y}) &= \psi(x\pm\hat{2})
\\
 \psi(x\pm\hat{z}) &= \psi(x\pm\hat{3})
\\ 
 \psi(x\pm\hat{\tau}) &= \psi(x\pm\hat{4}),
\end{split}
\end{equation}
where $\hat n$ is the unit vector in the $n$ direction.
We define the $z$-components of the vector current
\begin{equation}
\begin{split}
J^z =& -\frac{1}{2} \langle \bar\psi(x) (1-\gamma^3) \psi(x+\hat{3}) 
\\
&- \bar\psi(x+\hat{3}) (1+\gamma^3) \psi(x) \rangle 
\end{split}
\end{equation}
and the axial current
\begin{equation}
\begin{split}
J_5^z =& -\frac{1}{2} \langle \bar\psi(x) (1-\gamma^3) \gamma^5 \psi(x+\hat{3})
\\
&- \bar\psi(x+\hat{3}) (1+\gamma^3) \gamma^5 \psi(x) \rangle
.
\end{split}
\end{equation}
We computed these currents on the lattice with a dislocation.
The spatial lattice size is $12^3$.
At the boundaries in the $x$-$y$ plane, the derivative term perpendicular to the boundaries is set to be zero.
A periodic boundary condition is imposed in the $z$ direction and an anti-periodic periodic boundary condition is imposed in the $\tau$ direction.
The axial current has nonzero cutoff-dependent part even at $T=\mu=\mu_5=0$ (see Appendix \ref{secA}).
As shown in Fig.~\ref{figT}, the temperature dependence is in good agreement with a quadratic function as expected in Eq.~\eqref{eqJA}.
The cutoff-dependent part was estimated by the extrapolation $T \to 0$ at $\mu=\mu_5=0$, and then subtracted in Figs.~\ref{figJ3D} and \ref{figJ}.

\begin{figure}[t]
\begin{center}
 \includegraphics[width=.48\textwidth]{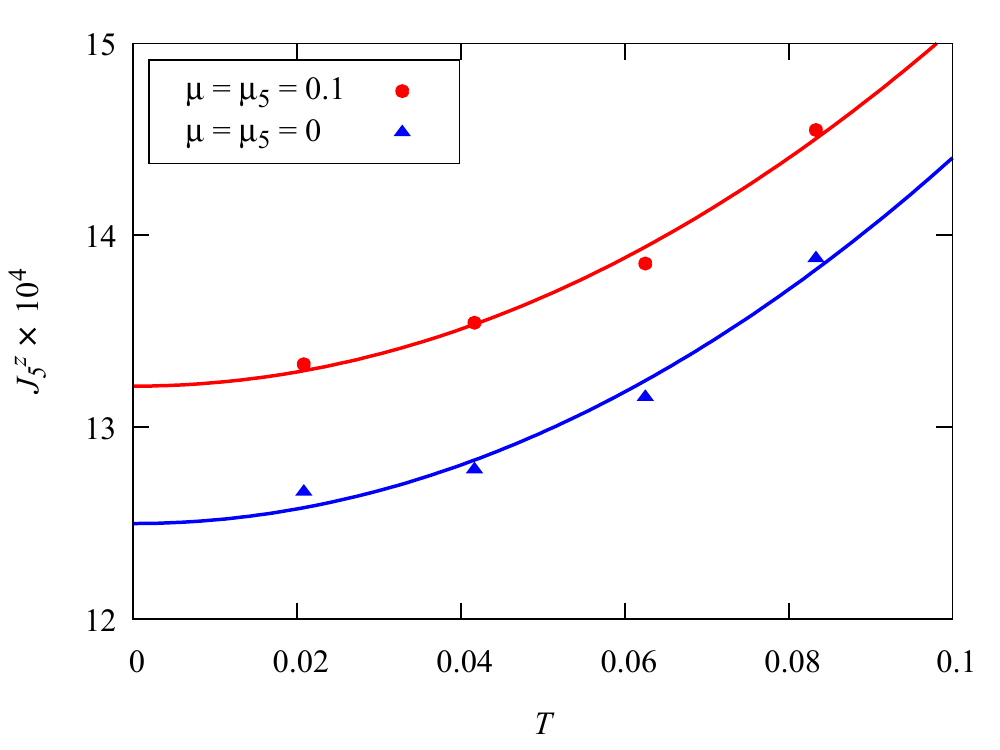}
\caption{
\label{figT}
Temperature dependence of the axial current $J_5^z$.
The Burgers vector length is $b=3$.
The solid curves are the best-fit quadratic functions.
The lattice unit is used.
}
\end{center}
\end{figure}

\section{Derivation of the chiral torsional effect}
\label{secA}

We analytically derive the chiral torsional effect.
The gravitational effect is expanded in the power series of the torsion tensor.
The metric tensor is assumed to be the flat one $g_{\mu\nu} = \delta_{\mu\nu}$ at each order.
For simplicity, we neglect the spin connection $\Gamma_\nu$ in the Dirac equation \eqref{eqDirac}.
The covariant derivative incorporates only the vector and axial chemical potentials,
$D_\nu \equiv \partial_\nu + \delta_{\nu \tau} (\mu + \gamma^5 \mu_5)$.
For later convenience, we also define
$\bar D_\nu \equiv \partial_\nu + \delta_{\nu \tau} (\mu - \gamma^5 \mu_5)$.

We compute the currents carried by right-handed and left-handed fermions,
\begin{align}
\begin{split}
	J_{\RIGHT/\LEFT}^m 
	&= \langle \bar{\psi} \gamma^m P_{\RIGHT/\LEFT} \psi \rangle
\\
	&= - \Tr_{\mathrm D, \mathrm C} \left( 
		\gamma^m P_{\RIGHT/\LEFT} \frac{1}{e_n^\nu \gamma^n D_\nu} 
	\right) ,
	\label{eq:JR}
\end{split}
\end{align}
where $P_{\RIGHT/\LEFT}=(1\pm\gamma^5)/2=(1\pm\gamma^1\gamma^2\gamma^3\gamma^4)/2$ is the chiral projector.
The trace is over the Dirac space ``$\mathrm D$'' as well as the coordinate space ``$\mathrm C$''.
Using the relation
\begin{align*}
	& (e^\mu_m \gamma^m D_\mu) (e^\nu_n \gamma^n \bar D_\nu)
	= \gamma^m \gamma^n (e^\mu_m \bar D_\mu) (e^\nu_n \bar D_\nu) \\
	&= \left( \frac12 \{\gamma^m, \gamma^n\} + \frac12 [\gamma^m, \gamma^n] \right)
	(e^\mu_m \bar D_\mu) (e^\nu_n \bar D_\nu) \\
	&= \bar D^2 - i \sigma^{mn} T^\nu_{mn} \bar D_\nu ,
\end{align*}
where $\sigma^{mn} \equiv (i/2) [\gamma^m, \gamma^n]$,
we expand the propagator in the power series of the torsion tensor.
\begin{align*}
	\frac{1}{e_n^\nu \gamma^n D_\nu} 
	& = e_n^\nu \gamma^n \bar D_\nu 
	\frac{1}{\bar D^2 - i \sigma^{rs} T^\alpha_{rs} \bar D_\alpha} \\
	& = e_n^\nu \gamma^n \bar D_\nu \left(
		\frac{1}{\bar D^2}
		+ i \frac{1}{\bar D^2} \sigma^{rs} T^\alpha_{rs} \bar D_\alpha \frac{1}{\bar D^2}
	+ \cdots \right).
\end{align*}
Plugging this expansion into the formula \eqref{eq:JR},
we get the zeroth and the linear orders of the currents.
For the right-handed sector,
\begin{align*}
	J_{\RIGHT}^m 
	= J_{\RIGHT\, (0)}^m + J_{\RIGHT\, (1)}^m + \cdots,
\end{align*}
where the zeroth order reads
\begin{align}
	J_{\RIGHT\, (0)}^m
	= - \Tr_{\mathrm D} (\gamma^m P_\RIGHT \gamma^n)
	\Tr_{\mathrm C} \left(
		e^\nu_n D_{\RIGHT \nu} \frac{1}{D_\RIGHT^2}
	\right)
	\label{J0}
\end{align}
and the linear order reads
\begin{align}
	&J_{\RIGHT\, (1)}^m
	= - i
	\Tr_{\mathrm D} (\gamma^m P_\RIGHT \gamma^n \sigma^{rs})
	\Tr_{\mathrm C} \left(
		e^\nu_n D_{\RIGHT \nu} \frac{1}{D_\RIGHT^2}
		T^\alpha_{rs} D_{\RIGHT \alpha} \frac{1}{D_\RIGHT^2}
	\right) .
	\label{J1}
\end{align}
We defined the covariant derivative for the right-handed spinor
$D_{\RIGHT \nu} = \partial_\nu + \delta_{\nu \tau} \mu_\RIGHT$
with $\mu_\RIGHT = \mu + \mu_5$.
We will evaluate each order in turn.

The Dirac trace in the zeroth order current \eqref{J0} reads
\begin{align}
	\Tr_{\mathrm D} (\gamma^m P_\RIGHT \gamma^n)
	= 2 \delta^{mn} .
	\label{J0D}
\end{align}
The trace over the coordinate space is given by the four-momentum integral
using the Fermi distribution function $n(x) = 1 / (e^{x/T} + 1)$.
\begin{align*}
	\Tr_{\mathrm C}
	\left( e^\nu_n D_{\RIGHT \nu} \frac{1}{D_\RIGHT^2} \right)
	= e^\tau_n \int_{\bv{k}} \int_C \frac{d \omega}{2\pi} \;
	\frac{-i K_{\RIGHT \tau}}{- K_\RIGHT^2} n(i \omega) ,
\end{align*}
where $K_{\RIGHT \nu} = (\bv{k}, \omega + i\mu_\RIGHT)$.
We neglected the momentum dependence of $e^\nu_n$ because it comes from the higher orders of the derivative of $e^\nu_n$.
The integral with respect to $\omega$,
which is a complex integral along the contour $C$
surrounding the Matsubara frequencies,
is evaluated by deforming the contour.
The remaining momentum integral,
$\int_{\bv{k}} \equiv \int d^3\bv{k} / (2\pi)^3$,
can also be carried out by introducing the cutoff at 
$k \equiv |\bv{k}| = \Lambda$.
The result reads
\begin{align}
\begin{split}
	& - e^\tau_n \int \frac{k^2 dk}{4 \pi^2} \;
	\left[
		n(k - \mu_\RIGHT) - n(k + \mu_\RIGHT) - 1
	\right] \\
	&= - e^\tau_n
	\left( 
	\frac{\mu_\RIGHT^3}{12 \pi^2} + \frac{\mu_\RIGHT T^2}{12}
	\right) 
	+ (\text{$\Lambda$-dependent}).
	\label{J0C}
\end{split}
\end{align}
Plugging the traces \eqref{J0D} and \eqref{J0C} into Eq.~\eqref{J0}, we get
\begin{align*}
	J_{\RIGHT/\LEFT\, (0)}^\mu
	= \delta^{\mu \tau}
	\left( \frac{\mu_{\RIGHT/\LEFT}^3}{6 \pi^2} + \frac{\mu_{\RIGHT/\LEFT} T^2}{6} \right)
	+ (\text{$\Lambda$-dependent}) .
\end{align*}
The left-handed current was obtained in the same way.
The difference $J_{5\, (0)}^\mu = J_{\RIGHT\, (0)}^\mu - J_{\LEFT\, (0)}^\mu$,
after the subtraction of the ultraviolet divergence,
gives the zeroth-order terms of the chiral imbalance \eqref{eqnA}.

Let us now evaluate the current at the linear order \eqref{J1}.
The Dirac trace reads
\begin{align}
	\Tr_{\mathrm D} (\gamma^m P_{\RIGHT/\LEFT} \gamma^n \sigma^{rs})
	= - 2i (\pm \epsilon^{mnrs} + \delta^{mr} \delta^{ns} - \delta^{ms} \delta^{nr}) .
	\label{J1D}
\end{align}
The trace over the coordinate space is given by
\begin{align*}
	&\Tr_{\mathrm C} \left(
		e^\nu_n D_{\RIGHT \nu} \frac{1}{D_\RIGHT^2}
		T^\alpha_{rs} D_{\RIGHT \alpha} \frac{1}{D_\RIGHT^2}
	\right) \\
	&= - e^\nu_n T^\alpha_{rs}
	\int_C \frac{d \omega}{2\pi} \int_{\bv{k}}
	\frac{K_{\RIGHT \nu} K_{\RIGHT \alpha}}{(K_\RIGHT^2)^2} n(i \omega) .
\end{align*}
Because of the rotational symmetry, this integral only depends on
the two tensors, 
$\delta^{\mathrm T}_{\nu\alpha}
\equiv \delta_{\nu \tau} \delta_{\alpha \tau}$ and 
$\delta^{\mathrm S}_{\nu\alpha}
\equiv \delta_{\nu\alpha} - \delta^{\mathrm T}_{\nu\alpha}$.
That is, 
\begin{align*}
	& \Tr_{\mathrm C} \left(
		e^\nu_n D_{\RIGHT \nu} \frac{1}{D_\RIGHT^2}
		T^\alpha_{rs} D_{\RIGHT \alpha} \frac{1}{D_\RIGHT^2}
	\right) \\
	&= - \delta^{\mathrm T}_{\nu\alpha} e^\nu_n T^\alpha_{rs} 
	\int_C \frac{d \omega}{2\pi} \int_{\bv{k}}
	\frac{(\omega + i \mu_\RIGHT)^2}{[(\omega + i\mu_\RIGHT)^2 + \bv{k}^2]^2} n(i \omega) \\
	&\quad
	- \delta^{\mathrm S}_{\nu\alpha} e^\nu_n T^\alpha_{rs} 
	\int_C \frac{d \omega}{2\pi} \int_{\bv{k}}
\frac{\bv{k}^2 / 3}{[(\omega + i\mu_\RIGHT)^2 + \bv{k}^2]^2} n(i \omega) .
\end{align*}
The integral in the first term is,
by carrying out the partial integration,
evaluated as
\begin{align*}
	& \int_C \frac{d \omega}{2\pi} \int \frac{dk}{4\pi^2} \;
	\frac{(\omega + i\mu_\RIGHT)^2}{(\omega + i\mu_\RIGHT)^2 + k^2} n(i \omega) \\
	&= - \int \frac{k dk}{8 \pi^2}
	\left[
		n(k - \mu_\RIGHT) + n(k + \mu_\RIGHT) - 1
	\right] \\
	&= - \left( \frac{ \mu_\RIGHT^2}{16 \pi^2} + \frac{T^2}{48} \right)
	+ (\text{$\Lambda$-dependent}) .
\end{align*}
The integral in the second term reads
\begin{align*}
	& \int_C \frac{d \omega}{2\pi} \int_{\bv{k}}
\frac{\bv{k}^2 / 3}{[(\omega + i\mu_\RIGHT)^2 + \bv{k}^2]^2} n(i \omega) \\
	&= \int_C \frac{d \omega}{2\pi} \int \frac{dk}{4\pi^2} \;
	\frac{k^2}{(\omega + i\mu_\RIGHT)^2 + k^2} n(i \omega) \\
	&= - \int_C \frac{d \omega}{2\pi} \int \frac{dk}{4\pi^2} \;
	\frac{(\omega + i\mu_\RIGHT)^2}{(\omega + i\mu_\RIGHT)^2 + k^2} n(i \omega) \\
	&\quad + (\text{$\Lambda$-dependent}) \\
	&= \frac{\mu_\RIGHT^2}{16 \pi^2} + \frac{T^2}{48}
	+ (\text{$\Lambda$-dependent}) .
\end{align*}
Thus the trace over the coordinate space is given by
\begin{align}
	\begin{split}
	& \Tr_{\mathrm C} \left(
		e^\nu_n D_{\RIGHT \nu} \frac{1}{D_\RIGHT^2}
		T^\alpha_{rs} D_{\RIGHT \alpha} \frac{1}{D_\RIGHT^2}
	\right) \\
	&= (\delta^{\mathrm T}_{\nu\alpha} - \delta^{\mathrm S}_{\nu\alpha}) 
	e^\nu_n T^\alpha_{rs} 
	\left( \frac{\mu_\RIGHT^2}{16 \pi^2} + \frac{T^2}{48} \right)
	+ (\text{$\Lambda$-dependent}) .
	\label{J1C}
	\end{split}
\end{align}
Plugging the traces \eqref{J1D} and \eqref{J1C} into Eq.~\eqref{J1},
we get the torsion-induced current.
Below the results are shown together with the left-handed one.
\begin{align*}
	\begin{split}
	J_{\RIGHT/\LEFT\, (1)}^\mu
	&= - (\delta^{\mathrm T}_{\nu\alpha} - \delta^{\mathrm S}_{\nu\alpha}) 
	(\pm \epsilon^{\mu\nu\rho\sigma} + 2 \delta^{\mu\rho} \delta^{\nu\sigma}) T^\alpha_{\rho\sigma} \\
	&\quad \cdot
	\left( \frac{\mu_{\RIGHT/\LEFT}^2}{8 \pi^2} + \frac{T^2}{24} \right)
	+ (\text{$\Lambda$-dependent}) .
	\end{split}
\end{align*}
The positive and negative signs associate to the right-handed and left-handed currents, respectively.
These currents are rephrased as the vector and axial currents,
which read
\begin{align}
	\begin{split}
		J_{(1)}^\mu &= J_{\RIGHT\, (1)}^\mu + J_{\LEFT\, (1)}^\mu \\
	&= - (\delta^{\mathrm T}_{\nu\alpha} - \delta^{\mathrm S}_{\nu\alpha}) T^\alpha_{\rho\sigma} \\
	&\quad \cdot
	\left[
		\epsilon^{\mu\nu\rho\sigma} \frac{\mu \mu_5}{2\pi^2}
		+ 2 \delta^{\mu\rho} \delta^{\nu\sigma} \left( \frac{\mu^2 + \mu_5^2}{4\pi^2} + \frac{T^2}{12} \right)
	\right]
	\label{JV}
	\end{split}
\end{align}
and
\begin{align}
	\begin{split}
		J_{5\, (1)}^\mu &= J_{\RIGHT\, (1)}^\mu - J_{\LEFT\, (1)}^\mu \\
	&= - (\delta^{\mathrm T}_{\nu\alpha} - \delta^{\mathrm S}_{\nu\alpha}) T^\alpha_{\rho\sigma} \\
	&\quad \cdot
	\left[
		\epsilon^{\mu\nu\rho\sigma} \left( \frac{\mu^2 + \mu_5^2}{4\pi^2} + \frac{T^2}{12} \right)
		+ 2 \delta^{\mu\rho} \delta^{\nu\sigma}  \frac{\mu \mu_5}{2\pi^2}
	\right] 
	\label{JA}
	\end{split}
\end{align}
after the subtraction of the ultraviolet divergence.
They give Eq.~\eqref{eqJV}, \eqref{eqJA}, \eqref{eqJVe}, and \eqref{eqJAe}.

In the above derivation, we have assumed that the spin connection is zero, $\Gamma_\mu =0$.
In general, Dirac fermions can couple to torsion through the spin connection.
The minimal one~\cite{Audretsch:1981xn} is
\begin{eqnarray} 
 \Gamma_\mu &=& \frac{1}{2} \omega_{\mu mn} \sigma^{mn}
\\
 \omega_{\mu mn} &=& \frac{1}{2} K^\gamma_{\mu\alpha} \delta_{\beta\gamma} e_m^\alpha e_n^\beta
\\
 K^\gamma_{\mu\alpha} &=& T^\gamma_{\mu\alpha} - T^\mu_{\alpha\gamma} + T^\alpha_{\gamma\mu}
.
\end{eqnarray}
This will give another contribution to $J_{(1)}^\mu$ and $J_{5\, (1)}^\mu$, though we do not take it into account in this paper.

\bibliography{paper}

\end{document}